\documentclass{cernyrep}
\usepackage[T1]{fontenc}
\usepackage[bookmarks, colorlinks=true, linktoc=page, pdftex, linkcolor=black, citecolor=black, urlcolor=blue]{hyperref}

\usepackage{fancyhdr}
\usepackage{subfig}
\usepackage[labelfont=bf]{caption}
\DeclareCaptionFont{tenpt}{\fontsize{10}{12}\selectfont}
\usepackage{url}

\fancyhfoffset{4 mm}
\fancypagestyle{ARTTITLE}{%
\fancyhf{} 
\lhead{\hfill CERN Accelerator School Proceedings --
{\it Intensity Limitations in Hadron Beams} --  Borovets, Bulgaria, 2025\hfill}
\lfoot{\hspace{3mm} Available online at \url{https://cas.web.cern.ch/previous-schools}}
\rfoot{\thepage\hspace*{3mm}}
 
}

\usepackage{varwidth}
\usepackage{xcolor}
\begin{document}

\title{Particle-Matter Interactions}
\author{Giuseppe Lerner}
\institute{CERN, Geneva, Switzerland}

\begin{abstract}
This lecture reviews the principles of particle–matter interactions, providing the essential physics background required to understand beam loss mechanisms in high-energy accelerators and their associated implications. The main interaction processes of photons and charged particles are introduced, together with an overview of nuclear reactions. The lecture then addresses electromagnetic and hadronic showers, which play a central role in particle-matter interaction physics. Following a brief overview of Monte Carlo simulation tools, with emphasis on FLUKA, the lecture concludes with a detailed examination of a representative LHC-type radiation shower.
\end{abstract}
\keywords{Particles; matter; interaction; radiation; physics; Monte Carlo; simulations.}
\maketitle
\thispagestyle{ARTTITLE}


\section{Introduction}

This article constitutes the first part of a two-lecture series on the topic of particle-matter interactions and beam loss consequences~\cite{bib:LernerCAS2025beamlossconsequences} at high-energy accelerators, covering their fundamental mechanisms and the~challenges that they pose at existing facilities, such as the Large Hadron Collider (LHC)\cite{bib:LHC}, and at future accelerators. After a brief digression on the meaning of terms such as radiation and particles (Section~\ref{sec:beams_particles}), the lecture begins with a review of the basic principles of particle-matter interactions in Section~\ref{sec:basic_concepts}. This is followed by a description of the interaction mechanisms of photons and charged particles with atoms (Section~\ref{sec:photons_charged_interactions}) and of nuclear interactions (Section~\ref{sec:nuclear_interactions}), providing the foundation for introducing electromagnetic and hadronic showers (Section~\ref{sec:EM_had_showers}). The article then describes Monte Carlo techniques to simulate radiation-matter interactions (Section~\ref{sec:MC_simulations}) before presenting a close analysis of a typical radiation shower initiated by the loss of a beam-like particle at the CERN Large Hadron Collider (LHC) in Section~\ref{sec:LHC_showers}.

\section{Particles or radiation}
\label{sec:beams_particles}

In the context of high-energy accelerators, the terms ``particle–matter interactions'' and ``radiation–matter interactions'' are often used interchangeably, sometimes alongside ``beam–matter interactions''. An accelerator physicist, accustomed to treating the beam as a collective entity, will likely prefer the latter, implying that the impact of any beam (or a fraction thereof) on material elements can trigger a chain of events with non-trivial consequences. A particle physicist would more commonly use ``particle–matter interactions'', a term that applies to any particle, regardless of whether it belongs to a primary beam or has a different origin. Finally, an engineer responsible for accelerator equipment, or a~safety officer concerned with facility accessibility, might use the term ``radiation'', emphasizing the key underlying issue, often with serious practical implications.

It turns out that the terms ``beam'', ``particle", and ``radiation'' are all correctly used in this scope, according to their meaning. In Quantum Field Theory (QFT), the fundamental science that governs the~phenomena under study, an elementary particle is defined as the quantum excitation of a field in a Lagrangian, which can be either truly elementary or composite, i.e., a bound state of multiple elementary excitations. In contrast, the concept of radiation refers to the transport of energy in space(time), carried by particles or waves in a quantum-mechanical sense. Broadly speaking, any wave that propagates in space represents a type of radiation, but in the context of high-energy accelerators it is often appropriate to emphasize the particle-like nature of the moving energy quanta. In other words, particles propagating through space (including those forming a circulating beam in an accelerator) \textit{are} radiation. When they interact with matter, they can simultaneously deposit energy, trigger secondary processes, and cause damage, as this lecture and the next will summarize. 

It is worth noting that radiation is sometimes mistakenly referred to as radioactivity. In reality, radioactivity is not the radiation itself but the nuclear decay process that produces it. While radioactivity is indeed relevant at accelerator facilities (particularly because of the exposure risks to personnel in activated areas accessed after operation), it is important to distinguish clearly between these two concepts.

\section{Basic elements}
\label{sec:basic_concepts}

\subsection{Particles and matter}

The list of elementary particles in the Standard Model is presented in Table~\ref{tab:SM_particles}. When dealing with particle-matter interactions at accelerators, an important role is played by hadrons, i.e., bound states of quarks, that are either stable or meta-stable, such as protons or neutrons, or have a long-enough lifetime to travel significant distances before decaying, such as charged pions or kaons. Neutral pions and kaons are also highly relevant, as will be further detailed. Among leptons, electrons are clearly an essential actor in radiation-matter interaction processes, while muons also play an important role. The role of photons as the mediator of electromagnetic interactions is also critical, while the other gauge bosons, the Higgs boson and neutrinos will not be referenced often in this lecture. Concerning matter, one may consider any material composed of elements from the well-known Mendeleev periodic table, or combinations thereof; moreover, the table of nuclides~\cite{bib:KAERI_nuchart} illustrates that each element can exist in multiple stable and unstable isotopic forms.

\begin{table}[ht]
\centering
\renewcommand{\arraystretch}{1.2}
\caption{Standard Model particles grouped by type.}
\begin{tabular}{lll}
\hline\hline
\textbf{Category} & \textbf{Subgroup} & \textbf{Particles} \\
\hline
Quarks &
Up-type &
$u,\ c,\ t$ \\
&
Down-type &
$d,\ s,\ b$ \\
\hline
Leptons &
Charged &
$e,\ \mu,\ \tau$ \\
&
Neutral (Neutrinos) &
$\nu_e,\ \nu_\mu,\ \nu_\tau$ \\
\hline
Gauge Bosons &
Force Carriers &
$\gamma,\ W^{\pm},\ Z,\ g$ \\
\hline
Scalar Boson &
Higgs &
$H$ \\
\hline\hline
\end{tabular}
\label{tab:SM_particles}
\end{table}


\subsection{Cross section and mean free path}

The interactions between particles, and by extension the interactions of particles with matter, are stochastic processes governed by the laws of quantum physics and relativity (i.e., Quantum Field Theory, as introduced above). A fundamental quantity describing particle interaction processes is the \textbf{cross section} $\sigma$, measured in barn ($1$~b~$=10^{-24}$~cm$^2$). In a counting experiment measuring the interaction rate $dN_{\mathrm{int}}/dt$ of incoming particles on a target, the cross section is defined operationally as: 
\begin{equation}
\label{eqn:cross_section}
    \frac{1}{V} \times \frac{dN_{\mathrm{int}}}{dt} = \sigma \times n_T \times \Phi,
\end{equation}
where $\Phi$ is the incoming particle flux on a volume $V$, and $n_T$ is the number density of a target: 
\begin{equation}
\label{eqn:n_T_def}
    n_T = \frac{\rho \times N_A}{M}
\end{equation}
with $\rho$ and $M$ being the mass density and the molar mass of the material, while $N_A = 6.022\cdot 10^{23}$~mol~$^{-1}$ is Avogadro's number. Crucially, the cross section defined in Eq.~(\ref{eqn:cross_section}) is a property of the physical process, depending exclusively on the identity and energy of the particles and on the target material, while the~remaining parameters in the equation merely serve as normalization factors.

\begin{figure}[ht]
    \centering 
    \subfloat[\label{fig:thickness_material_scheme}]{
    \includegraphics[width=0.35\linewidth]{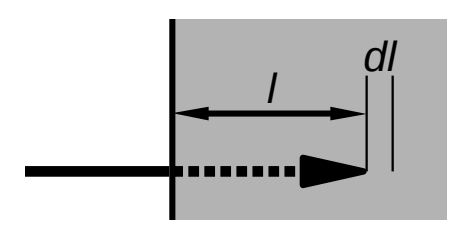}} \hspace{2cm}
    \subfloat[\label{fig:table_elements}]{
    \includegraphics[width=0.2\linewidth]{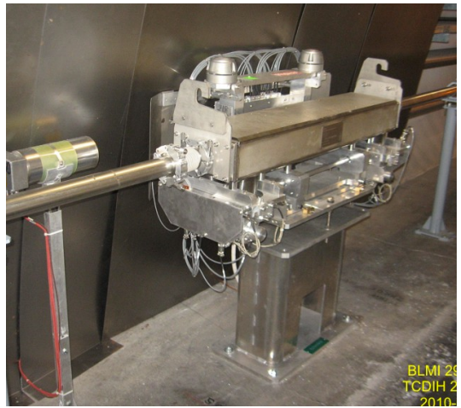}} 
    \caption{(a) Particles incident on a material at a depth $l$, and (b) photo of a TCDIH collimator in the transfer line from the SPS to the LHC at CERN.}
    \label{fig:interaction_probability_example}
\end{figure}

From the cross section one can directly compute the mean free path of particles in a material: 
\begin{equation}
\label{eqn:lambda_def}
    \lambda = \frac{1}{n_T \times \sigma}.
\end{equation}
In turn, this makes it possible to compute the integral survival probability $P_s$ of a beam of N particles incident on a homogeneous target as a function of the depth $l$, as illustrated in Fig.~\ref{fig:thickness_material_scheme}: 
\begin{equation}
    \frac{dN}{N} = - \frac{dl}{\lambda} \; \Rightarrow \; \int \frac{dN}{N} = - \frac{l}{\lambda} \; \Rightarrow \; \ln{\left[ \frac{N(l)}{N_0} \right]} = - \frac{l}{\lambda} \; \Rightarrow \; P_s (l) = \frac{N(l)}{N_0} = e^{-l/\lambda}
\end{equation}
and the closely related interaction probability $P_{\mathrm{int}}$: 
\begin{equation}
\label{eqn:interaction_probability}
    P_{\mathrm{int}} = 1 - P_s(l) = 1-e^{-l/\lambda},
\end{equation}
which can be approximated as $P_{\mathrm{int}} \approx l/\lambda$ for thin targets ($l << \lambda$).

\subsection{An example at the LHC}
\label{par:LHC_450GeV_example}

As an example, one can consider the case of a bunch of $450$~GeV protons (with $1.6\cdot 10^{11}$~p/bunch) injected from the Super Proton Synchrotron (SPS) into the LHC, and assume that they are all intercepted (by accident) by a $1.2$~m-long Graphite collimator (with density $\rho = 1.84$~g/cm$^3$). Using the inelastic interaction cross section of such $450$~GeV protons on Carbon at rest ($\sigma\approx 245$~mb), one can compute the~proton mean free path in the collimator by combining Eqs.~(\ref{eqn:n_T_def})~and~(\ref{eqn:lambda_def}): 
\begin{equation}
    \lambda = \frac{12~\mathrm{g/mol}}{1.84~\mathrm{g/cm}^3\times 6.022 \cdot 10^{23}~\mathrm{mol}^{-1} \times 245 \cdot 10^{-27}~\mathrm{cm}^2} \approx 44~\mathrm{cm}.
\end{equation}
From this, one can then derive the number of inelastic collisions, i.e., the number of lost protons in the~collimator, using Eq.~(\ref{eqn:interaction_probability}):
\begin{equation}
    N_{\mathrm{int}} = N_{\mathrm{p/bunch}} \times P_{\mathrm{int}} = 1.6 \cdot 10^{11}~\mathrm{p} \times \left[ 1-e^{-(120~\mathrm{cm}/44~\mathrm{cm})} \right] \approx 1.5\cdot10^{11}~\mathrm{p},
\end{equation}
which means that the vast majority of protons are lost locally. 



\section{Photon and charged particle interactions with atoms}
\label{sec:photons_charged_interactions}

\subsection{Photon interactions}

\begin{figure}[ht]
    \centering 
    \subfloat[\label{fig:photon_interaction_1a}]{
    \includegraphics[width=0.32\linewidth]{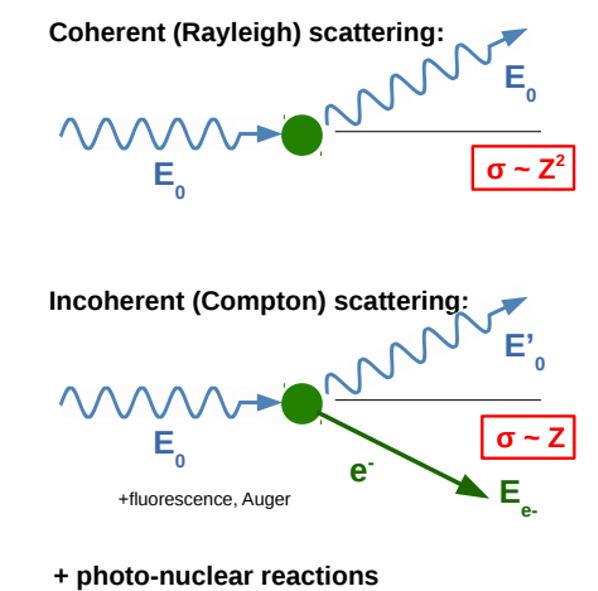}}
    \subfloat[\label{fig:photon_interaction_1b}]{
    \includegraphics[width=0.313\linewidth]{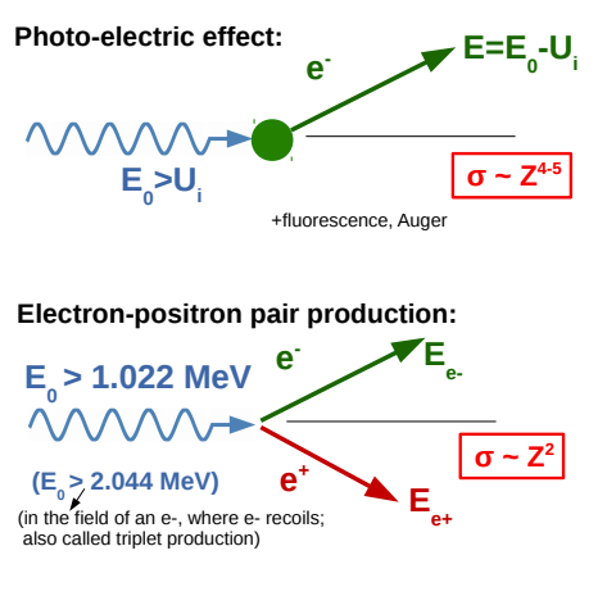}} 
    \subfloat[\label{fig:photon_interaction_cross_section}]{
    \includegraphics[width=0.34\linewidth]{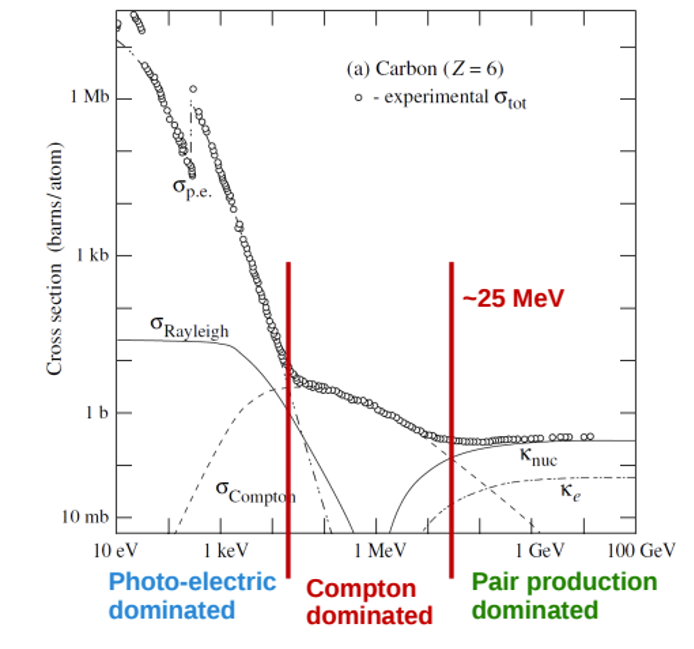}} 
    \caption{Photon interaction mechanisms, showing the diagrams of (a) Rayleigh and Compton scattering and (b)~photoelectric effect and electron-positron pair production (from Ref.~\cite{bib:Penelope2018}), as well as (c) the photon interaction cross section as a function of energy in Carbon (from Ref.~\cite{bib:PDG2016}).}
    \label{fig:photon_interaction_1}
\end{figure}

Our review of the interaction mechanisms of different particles begins with photons, i.e., the massless mediators of the electromagnetic force. Multiple physics processes are capable of producing photons, among which we can cite Bremsstrahlung emission by charged particles and electron-positron annihilation in matter, particle decay (especially of neutral pions), and nuclear reactions such as radiative neutron capture or gamma de-excitations. Once produced, photons can interact with matter via the processes illustrated in Fig.~\ref{fig:photon_interaction_1}, along with a characteristic cross section curve versus energy in Carbon: 
\begin{itemize}
\item \textbf{Coherent (Rayleigh) scattering}: elastic scattering, most relevant at low energies where the wavelength is larger than the atomic size. The cross section scales approximately with the square of the~atomic number ($\sigma \propto$~Z$^2$).
    \item \textbf{Photoelectric effect}: absorption of a photon by a bound electron, which is then emitted. The cross section exhibits sharp edges linked to atomic shell binding energies, as visible just below the keV scale in Fig.~\ref{fig:photon_interaction_cross_section}, and it strongly depends on the atomic number ($\sigma \propto$~Z$^{4-5}$). As the process creates an atomic vacancy, it can lead to fluorescence (photon) and/or Auger (electron) emission.
    \item \textbf{Incoherent (Compton) scattering}: inelastic scattering from loosely bound electrons, producing a recoil electron and a scattered photon. Compton scattering is typically dominant at intermediate photon energies (roughly $100$~keV-$10$~MeV), and it does not show the sharp cross section edges of the photoelectric effect. The dependence of the cross section with the atomic number is linear ($\sigma \propto$~Z) and the process can be followed by fluorescence or Auger emission.  
    \item \textbf{Electron-positron pair production}: production of an electron-positron pair in the electromagnetic field of the atomic nucleus (with threshold $E_{\mathrm{th}} = 2\mathrm{m}_e = 1.022$~MeV) or, more rarely, in the~field of an electron (with $E_{\mathrm{th}} = 4\mathrm{m}_e = 2.044$~MeV). This is the dominant photon interaction mechanism at high energies (roughly above $10$~MeV), with a quadratic dependence on the atomic number ($\sigma \propto$~Z$^2$). 
    \item \textbf{Photonuclear reactions}: interactions where the photon is absorbed by the nucleus and the final state typically involves the emission of nucleons or heavier nuclear fragments. The cross section has a threshold typically around or just below $10$~MeV, a Giant Dipole Resonance (GDR) peak in the $15$-$25$~MeV range, an intermediate quasi-deuteron region extending up to $\mathcal{O}(100)$~MeV, and a~high-energy regime where the reaction mechanisms resemble those induced by charged hadrons.  
    
\end{itemize}

\begin{figure}[ht]
    \centering 
    \includegraphics[width=0.6\linewidth]{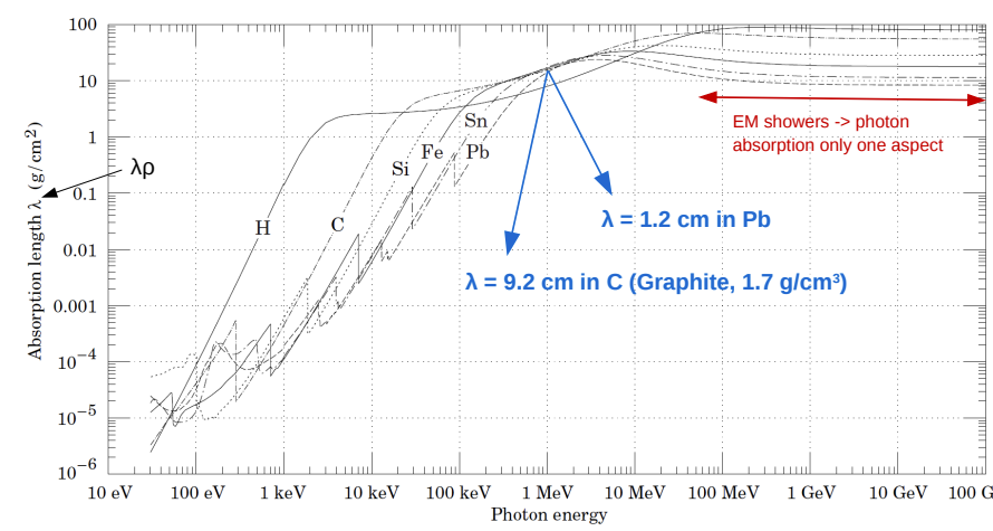}
    \caption{Mass attenuation length of photons as a function of energy in different materials (from Ref.~\cite{bib:PDG2016}). \label{fig:mass_attenuation_photons}}
\end{figure}

An important concept linked to photon interactions in material is the mass attenuation length $\lambda_\rho$, which is the characteristic length (multiplied by the mass density of the material) of the exponential law describing the photon intensity drop as a function of depth in a material: 
\begin{equation}
    I_{\gamma} (l) = I_{\gamma} (0) \times e^{-l\rho/\lambda _\rho} \quad \mathrm{where} \quad \lambda_\rho = \frac{M}{N_A \times \sigma_{\mathrm{tot}}}  
\end{equation}
with $M$ and $N_A$ defined as in Eq.~(\ref{eqn:n_T_def}). The mass attenuation length in different materials is shown in Fig.~\ref{fig:mass_attenuation_photons} as a function of photon energies, where the sharp increase from low energy to the MeV scale is linked to the corresponding drop in the cross section curve (Fig.~\ref{fig:photon_interaction_cross_section}), and the differences between materials are most evident below $\approx100$~keV, where the photoelectric effect is dominant. 

\subsection{Charged particle interactions}

\begin{figure}[ht]
    \centering 
    \subfloat[\label{fig:charged_particles_dE_dx}]{
    \includegraphics[width=0.535\linewidth]{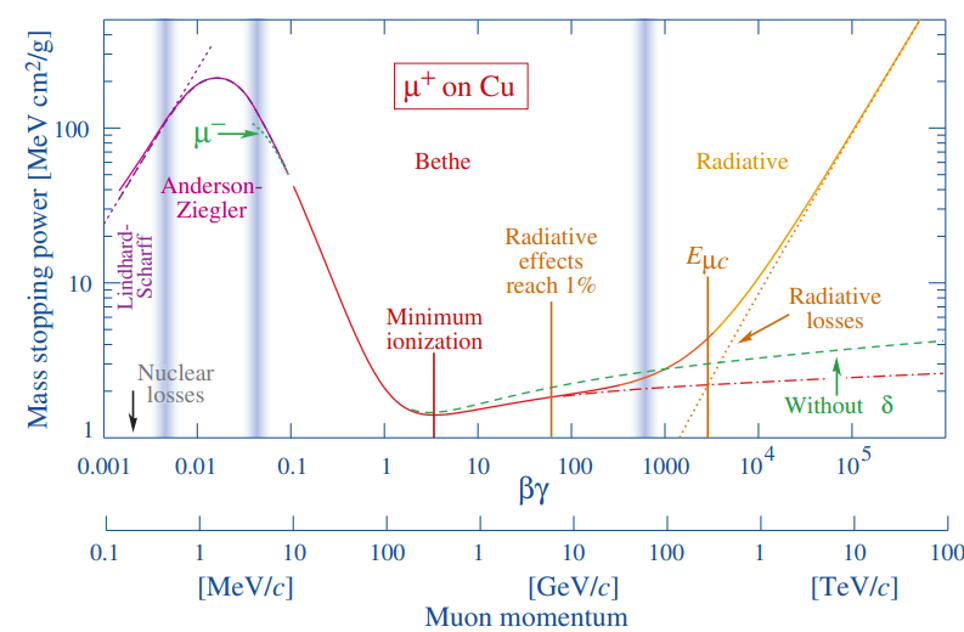}}
    \subfloat[\label{fig:charged_particles_critical_energy_vs_z}]{
    \includegraphics[width=0.445\linewidth]{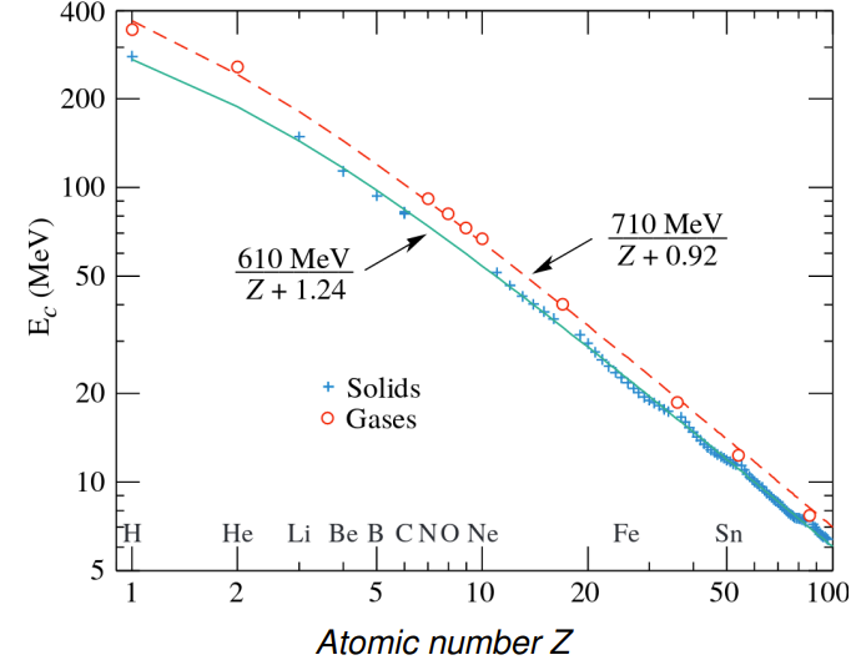}} 
    \caption{Graphs showing (a) the stopping power of positive muons in Copper and (b) the critical energy of electrons as a function of the atomic number of the target elements (from Ref.~\cite{bib:PDG2016}). }
    \label{fig:charged_particles_energy_loss}
\end{figure}

The interaction of charged particles with matter is primarily governed by the electromagnetic force, with the notable exception of high-energy charged hadrons, for which nuclear interactions can play a~significant role (see Section~\ref{sec:nuclear_interactions}). The associated phenomenology can be classified as follows:
\begin{itemize}
    \item \textbf{Coulomb interactions with atomic electrons}: they lead to the ionisation (or excitation) of the~atoms, with the production of knock-off electrons (known as $\delta$-rays if highly energetic) and leading to the progressive deposition of the projectile energy in the material. 
    \item \textbf{Coulomb interactions with atomic nuclei}: they cause Non-Ionising Energy Loss (NIEL), which is generally less relevant than atomic electron losses (except for low-energy heavy ions), but they are the dominant source of angular deflection for charged particles travelling through a material. 
    \item \textbf{Radiative processes}: they result from the interaction of the incoming projectile with the electromagnetic field of the atomic nuclei, leading to the emission of photons (referred to as Bremsstrahlung). 
\end{itemize}
The characteristic curve of the energy loss of charged particles as a function of particle momentum (or, more generally, of the relativistic $\beta\gamma$) is referred to as the \textbf{mass stopping power},
\begin{equation}
    -\frac{1}{\rho}\frac{dE}{dx}, \quad \mathrm{in~units~of}\left[ \mathrm{MeV}\cdot \mathrm{cm}^2\cdot \mathrm{g} ^{-1} \right],
\end{equation}
normalized to the density of the target material. Its overall shape is similar for most particles, as shown in Fig.~\ref{fig:charged_particles_dE_dx} for the case of positive muons in Copper. Electrons, however, show more pronounced differences due to their low mass and their identical nature to the atomic electrons with which they interact. Neglecting the lowest energies, where collective effects in the target material become important, the stopping power in Fig.~\ref{fig:charged_particles_dE_dx} decreases from its low-energy maximum approximately as $1/\beta^2$, or equivalently $1/(\beta\gamma)^2$, as described by the Bethe–Bloch formula. This behaviour arises from a kinematic enhancement of electromagnetic interactions at low projectile velocities, which can intuitively be interpreted as an increased interaction time with target electrons. At higher energies, relativistic effects progressively counterbalance this decrease, leading to a plateau known as the Minimum Ionising Particle (MIP) region. Eventually, for highly relativistic particles, radiative losses become dominant, leading to a sharp rise of the stopping power at high energies. 

A key concept in radiation-matter interaction physics is the \textbf{critical energy} $E_c$, defined as the~ener\-gy at which ionisation and radiative losses are equal. For most particles it is very large, typically of the~order of hundreds of GeV or more, while it is substantially lower for electrons due to their light mass. The critical energy of electrons is illustrated in Fig.~\ref{fig:charged_particles_critical_energy_vs_z} as a function of the atomic number, decreasing from several hundred MeV in light materials to below $10$~MeV for the heaviest elements.

\begin{figure}[ht]
    \centering 
    \includegraphics[width=0.6\linewidth]{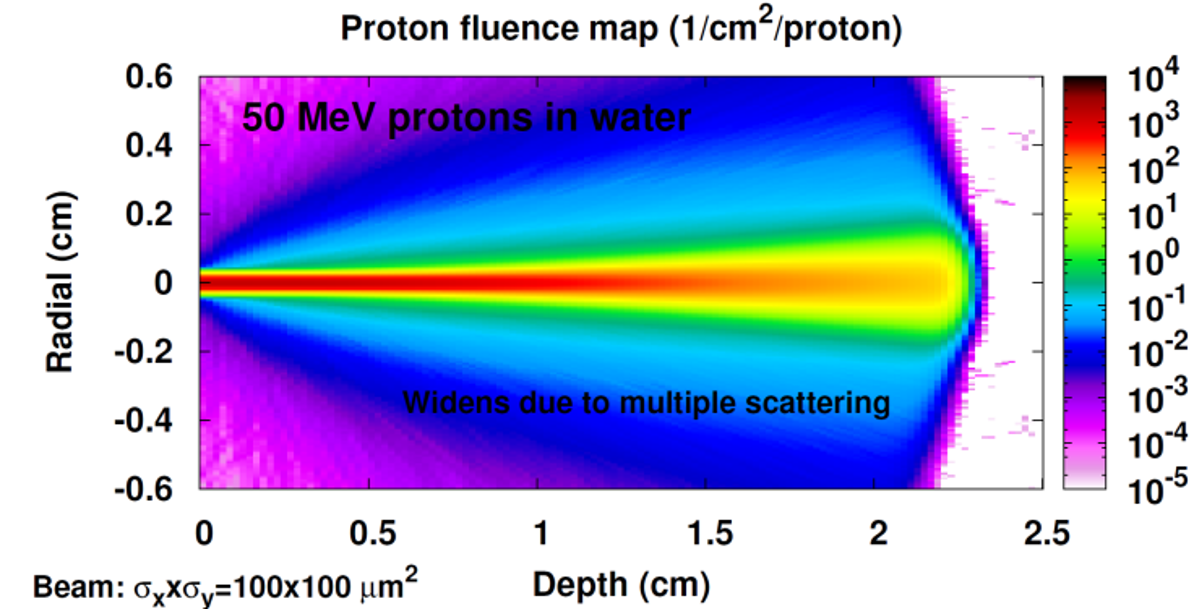}
    \caption{Proton fluence map in water for $50$ MeV protons simulated with FLUKA, where the broadening is caused by multiple Coulomb scattering. \label{fig:mcs_FLUKA}}
\end{figure}

Lastly, as discussed above, angular deflections of charged particles are primarily caused by successive Coulomb interactions with atomic nuclei and are well described by Molière’s theory of Multiple Coulomb Scattering (MCS)~\cite{bib:LynchDahl1991}. Without going into the full details, the key features of MCS are that scattering angles increase for lower-energy particles and for lighter particles at the same velocity. An~example of MCS is illustrated in Fig.~\ref{fig:mcs_FLUKA}, which shows the broadening of a 50-MeV proton beam in water, simulated with FLUKA over a depth of approximately 2~cm, before the protons come to rest due to ionisation losses.

\section{Nuclear interactions}
\label{sec:nuclear_interactions}

When hadrons are involved, interactions with atomic nuclei mediated by the strong force can occur and often dominate the resulting phenomenology, with weaker but sometimes relevant electroweak contributions. As a starting point, one can classify nuclear interactions as elastic or inelastic, depending on whether the internal structure of the particles is altered. In elastic interactions, the initial and final states of the interacting particles are the same, and the interaction only results in energy transfer (including the recoil of target nuclei in projectile–nucleus collisions) and angular deflections. These processes do not exhibit energy thresholds and are particularly relevant for neutral projectiles such as neutrons at low energies, while for charged projectiles elastic nuclear scattering can compete with Coulomb scattering on the~target nucleus. In contrast, inelastic interactions involve changes to the internal structure of the~particles and/or the production of new particles. Such processes generally exhibit energy thresholds, reflecting the minimum energy needed to disrupt nuclei or incoming particles, excite nuclear states, or create new particles. An important exception is neutron capture, which is threshold-free and has a cross section that peaks at thermal neutron energies (meV scale).

\begin{figure}[ht]
    \centering 
    \subfloat[\label{fig:nucleon_cross_sections}]{
    \includegraphics[width=0.525\linewidth]{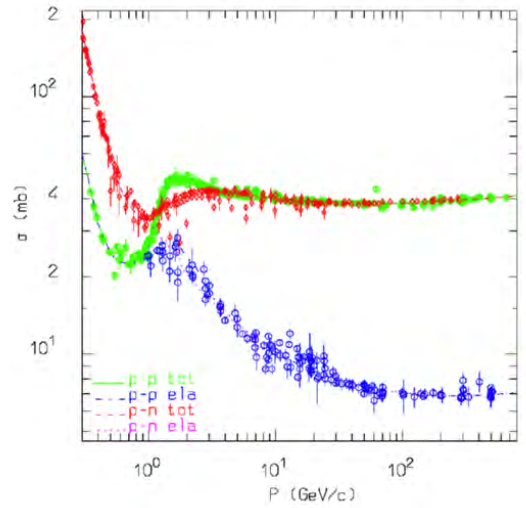}}
    \subfloat[\label{fig:neutron_cross_sections}]{
    \includegraphics[width=0.462\linewidth]{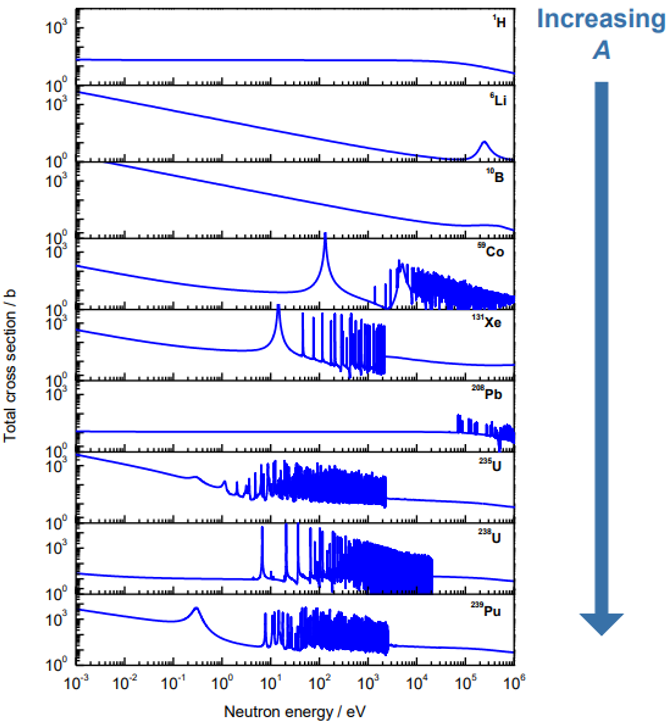}}
    \caption{(a) Nucleon-nucleon interaction cross sections (from Ref.~\cite{bib:MokhovCerutti2016}), and (b) neutron cross sections in different materials (from Ref.~\cite{bib:Schillebeeckx2014}). }
    \label{fig:nuclear_interactions}
\end{figure}

To understand the phenomenology of nuclear reactions, it is useful to first examine nucleon–nucleon collisions, for which the cross section is shown in Fig.~\ref{fig:nucleon_cross_sections} as a function of momentum. Elastic interactions dominate at low energies, while at intermediate energies the reactions proceed through intermediate resonant states (such as the $\Delta(1232)$ resonance or others). At high energies, the~cross section is governed by inelastic processes involving individual partons (i.e., the elementary constituents of the nucleon, such as quarks and gluons), and approaches an approximately asymptotic value. 

Concerning hadron-nucleus reactions, one can identify two characteristic stages. In the fast stage (lasting $\sim10^{-22}$~s), the projectile with energy $E$ interacts with nucleons, producing a large number of secondary particles (mostly pions, but also other hadrons and photons) with a multiplicity roughly proportional to $\log(E)$. The most energetic secondaries escape the nucleus and initiate a forward-directed hadronic cascade (further described in Section~\ref{sec:EM_had_showers}), while lower-energy secondaries deposit energy within the nucleus, leaving it in an excited state. In the subsequent slow stage ($\sim10^{-16}$~s), the excited nucleus relaxes by emitting MeV-scale light fragments (e.g., individual nucleons, or $\alpha$-particles) through evaporation and photons from $\gamma$ de-excitation, and heavier nuclei may also undergo fission when energetically allowed. Overall, the reaction yields an energetic hadronic shower from the fast stage (for sufficiently energetic projectiles), MeV-scale nuclear fragments and photons, and, optionally, fission products.

The case of neutron-induced reactions deserves special attention, as neutrons are hadrons with unique features. Their long half-life (over $10$ minutes) makes them effectively stable over the relevant timescales of radiation-matter interactions, and their lack of electric charge means they do not experience Coulomb forces, allowing them to travel long distances through matter. As they propagate, neutrons gradually lose energy through a series of elastic nuclear interactions (a process known as moderation) and they can eventually reach thermal energies (roughly $\approx25$~meV at room temperature). The neutron cross section varies strongly with energy and the target nucleus, as illustrated in Fig.~\ref{fig:neutron_cross_sections}. At low energies (below the eV scale), the cross section typically decreases with velocity ($\sigma \propto 1/v$), and neutron capture reactions become significant. In the intermediate range (eV–MeV), nuclear resonances appear, whose shapes and sizes have a complex dependence on the target nucleus and are difficult to model. Lastly, above a few MeV, neutron-induced interactions become more similar to those of other hadrons.

\section{Electromagnetic and hadronic showers and the Bragg peak}
\label{sec:EM_had_showers}

\begin{figure}[ht]
    \centering 
    \subfloat[\label{fig:EM_shower}]{
    \includegraphics[width=0.385\linewidth]{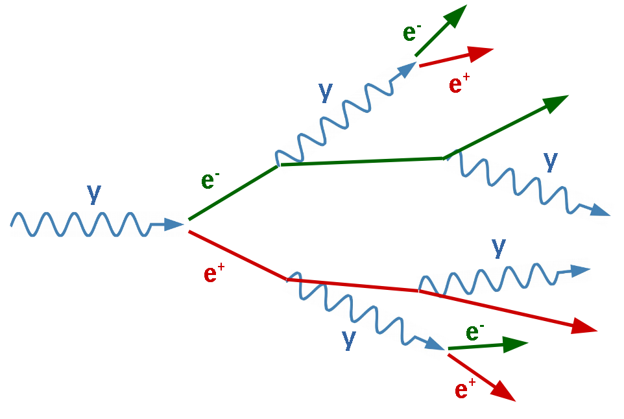}}
    \subfloat[\label{fig:had_shower}]{
    \includegraphics[width=0.594\linewidth]{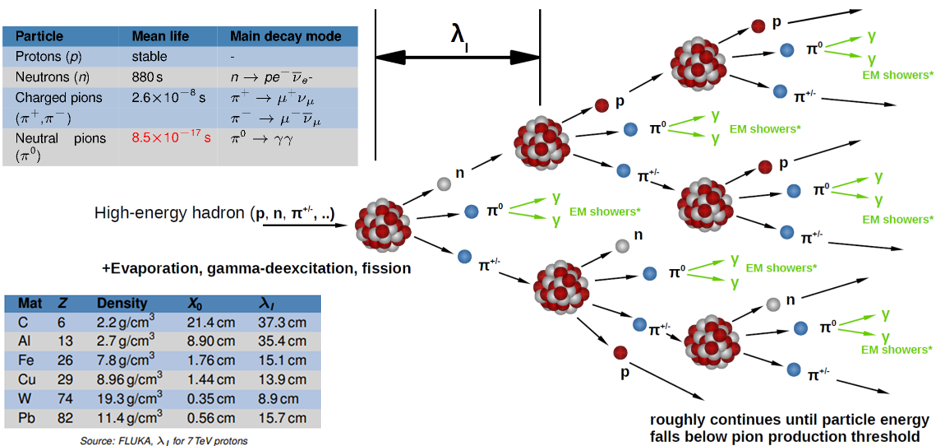}}
    \caption{Diagrams of (a) EM and (b) hadronic showers, with tables showing the main decay mode of relevant hadrons and the radiation and nuclear interaction lengths of reference materials. }
    \label{fig:EM_had_showers}
\end{figure}

As anticipated by the example in Paragraph~\ref{par:LHC_450GeV_example}, prompt radiation at high-energy accelerators is typically produced by interactions of beam particles with matter. If the projectile is an electron, positron, or photon, the interaction gives rise to an electromagnetic (EM) shower, shown in Fig.~\ref{fig:EM_shower}, consisting of a sequence of bremsstrahlung emissions by electrons and positrons and pair-production reactions by photons. In the case of hadronic projectiles, a sequence of inelastic nuclear reactions develops, producing a more complex cascade involving both hadronic and leptonic components, as illustrated in Fig.~\ref{fig:had_shower}. In both cases, provided the projectile energy is sufficiently high, the shower properties are largely independent of whether the initiating particle is an electron, positron, or photon (for EM showers), or of the specific hadron (for hadronic showers).

\begin{figure}[ht]
    \centering 
    \subfloat[\label{fig:EM_shower_multiplicity}]{
    \includegraphics[width=0.547\linewidth]{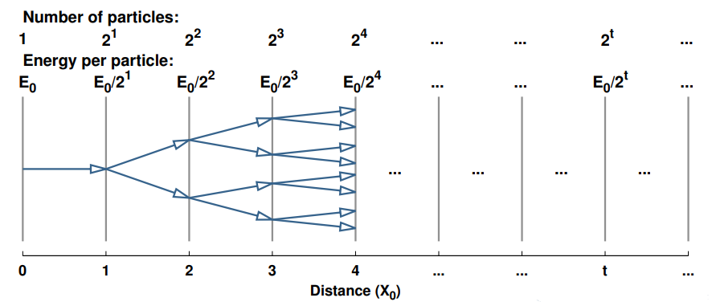}}
    \subfloat[\label{fig:EM_shower_edep_Copper}]{
    \includegraphics[width=0.438\linewidth]{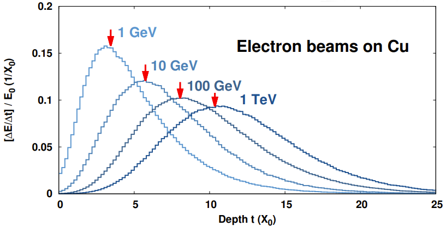}}
    \caption{(a) EM shower multiplicity as a function of depth and (b) energy deposition profile for EM showers initiated by electron beams with different energies on a Copper target. }
    \label{fig:EM_showers_propagation}
\end{figure}

For EM showers, the cascade develops as long as the individual particles remain above the critical energy $E_c$, defined in Section~\ref{sec:photons_charged_interactions} and shown in Fig.~\ref{fig:charged_particles_critical_energy_vs_z} as a function of the target atomic number. A~key material parameter governing this development is the \textbf{radiation length} $X_0$, corresponding to the~mean distance over which the energy of electrons or positrons is reduced to $1/e$ via Bremsstrahlung, or $7/9$ of the mean free path for electron/positron pair production by photons. The radiation length is inversely proportional to the material density and to the quantity $Z^2/A$ of the target, where $Z$ is the~atomic number and $A$ the atomic mass; values for different materials are summarized in table format in Fig.~\ref{fig:EM_had_showers}. In practice, the radiation length sets the longitudinal scale of EM showers in matter, as illustrated schematically in Fig.~\ref{fig:EM_shower_multiplicity}, where particle multiplicity is shown as a function of depth in units of $X_0$ (with $t \equiv z/X_0$). The~maximum multiplicity is reached at $t_\mathrm{max} \propto \ln (E_0 / E_c)$, typically a few radiation lengths depending on the projectile energy and material. The energy deposition profile, important to assess the impact of the shower on the target material, is shown in Fig.~\ref{fig:EM_shower_edep_Copper} for a Copper target and different electron energies. The~peak of the energy deposition curves is typically reached slightly upstream of $t_\mathrm{max}$, hence also scaling logarithmically with the projectile energy. 

The longitudinal scale of hadronic showers is also set by a well-defined parameter associated with the target material, known as the \textbf{nuclear interaction length} $\lambda_I$, which is listed in the table in Fig.~\ref{fig:EM_had_showers} along with $X_0$. It represents the mean free path between inelastic nuclear interactions and is approximately constant at high energies, reflecting the plateau of the corresponding cross sections (Fig.~\ref{fig:nucleon_cross_sections}). Hadronic showers have a richer particle content than EM showers. Multiple hadrons are produced in the~fast stage of nuclear reactions, accompanied by nuclear fragments and evaporation products. In addition, neutral pions decay almost immediately into photon pairs, generating secondary EM cascades within the main hadronic shower. The fraction of energy transferred to these EM particles, commonly called the EM fraction of the shower, is a stochastic quantity that depends on the relative number of neutral pions produced in the inelastic interactions. Overall, the hadronic shower continues as long as the secondary hadrons are above the production threshold of pions. Since $\lambda_I$ exceeds $X_0$ in all materials, and particularly in high-$Z$ ones, hadronic showers are typically longer than the corresponding EM ones, implying that hadronic radiation is typically harder to shield compared to EM radiation.

\begin{figure}[ht]
    \centering 
    \includegraphics[width=0.6\linewidth]{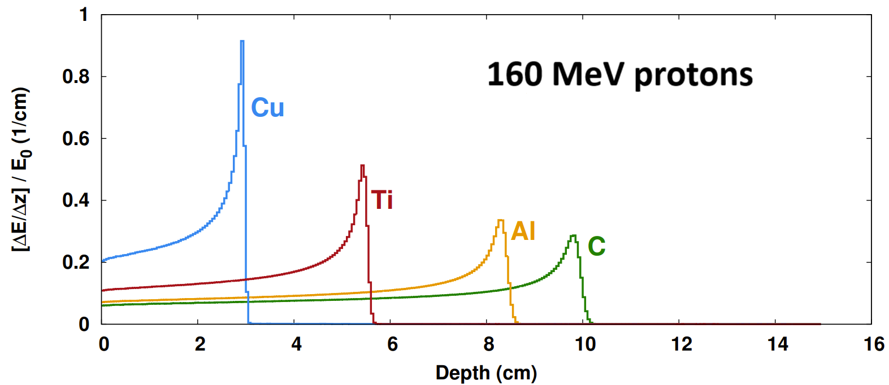}
    \caption{Energy deposition profile from $160$~MeV protons impacting on different materials, showing Bragg peaks at different depths. }
    \label{fig:Bragg_peak}
\end{figure}

If charged hadrons, such as protons, enter matter with energies below the pion production threshold (a few hundred MeV), they lose energy predominantly through ionisation and eventually come to rest in the material, as described by the stopping power shown in Fig.~\ref{fig:charged_particles_dE_dx}. Because the ionisation energy loss increases as the projectile slows down, the energy deposition rises toward the end of the particle’s range, leading to the characteristic Bragg peak shown in Fig.~\ref{fig:Bragg_peak} for the case of $160$~MeV protons. This feature is particularly relevant for medical physics applications such as proton therapy, where the Bragg peak is exploited to irradiate cancer in well-defined positions within the patient's body.

\section{Monte Carlo simulations of radiation-matter interaction}
\label{sec:MC_simulations}

Since particle-matter interactions are fundamentally stochastic, Monte Carlo (MC) codes provide an~optimal framework for simulating particle transport in matter. A wide range of tools is available for this purpose, including FLUKA~\cite{bib:FLUKA_website,bib:Ahdida2022,bib:Battistoni2015}, Geant4\cite{bib:Agostinelli2003,bib:Allison2016}, PHITS~\cite{bib:Sato2018}, MCNP~\cite{bib:Werner2018}, PENELOPE~\cite{bib:Penelope2018}, and others. Despite differences in scope and implementation, these codes share a similar conceptual structure: the user specifies a radiation source and a description of the geometry through which the~particles propagate, after which the code simulates their transport and computes a set of user-defined output observables. 

\begin{figure}[ht]
    \centering 
    \subfloat[\label{fig:FLUKA_tunnel_geometry}]{
    \includegraphics[width=0.6\linewidth]{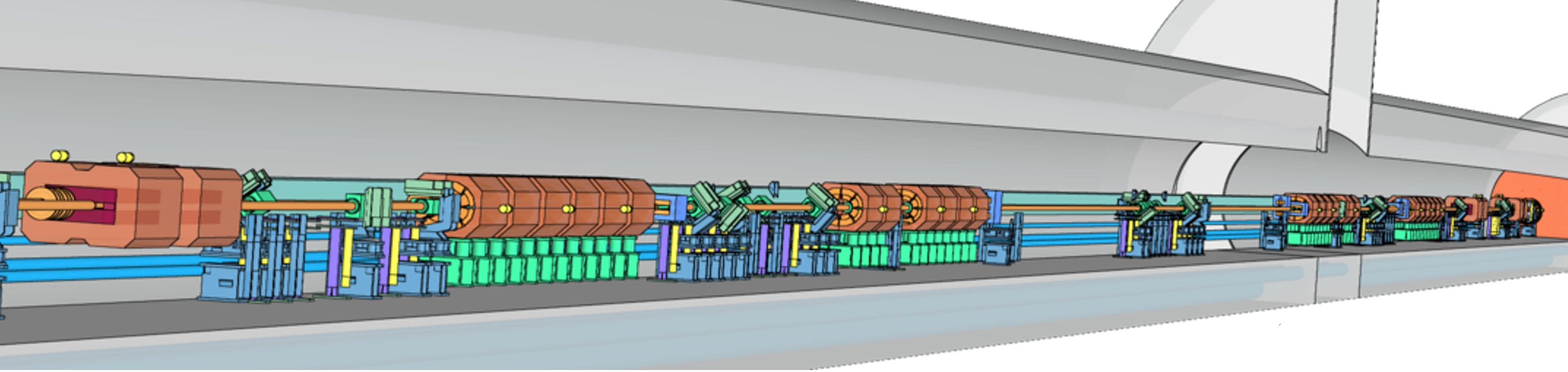}}
    \subfloat[\label{fig:FLUKA_random_path}]{
    \includegraphics[width=0.385\linewidth]{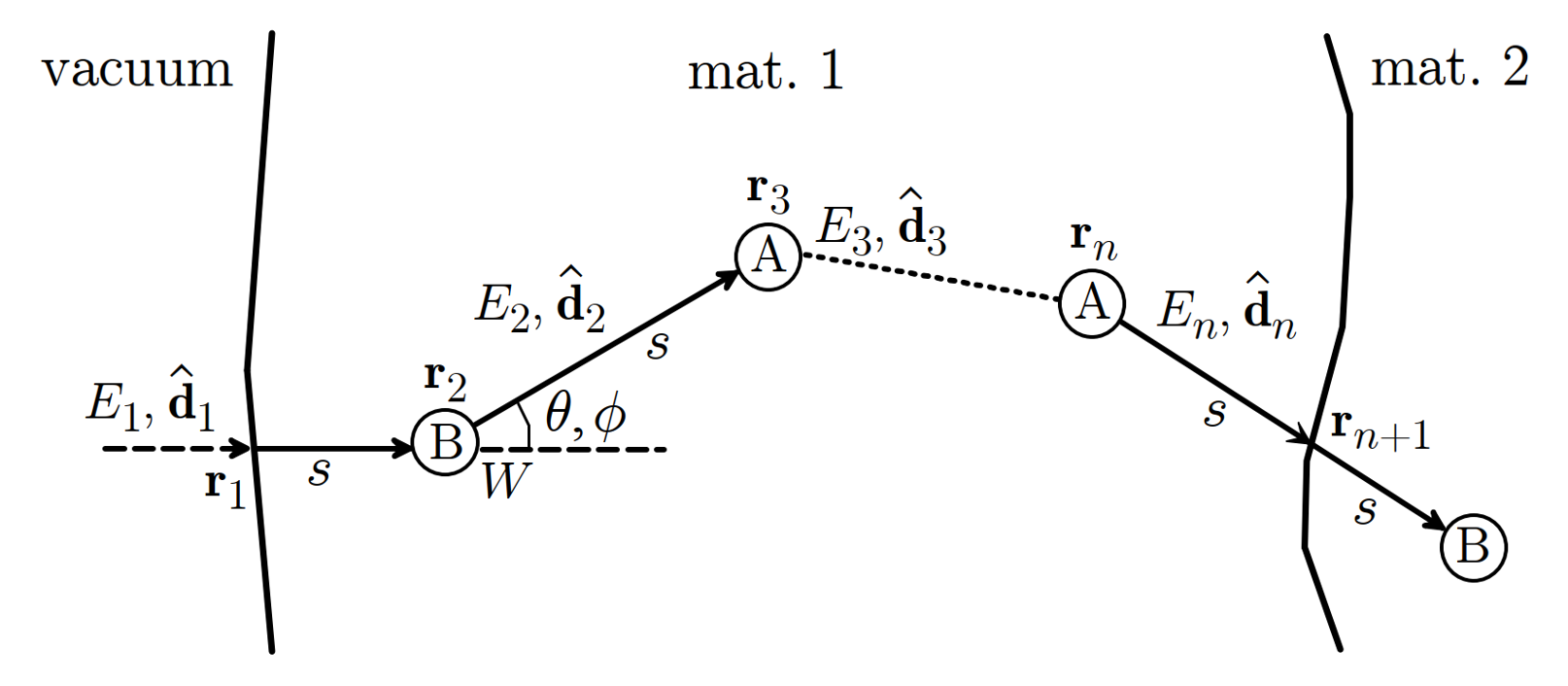}}
    \caption{(a) FLUKA geometry of a portion of LHC tunnel and (b) multi-step particle transport. }
    \label{FLUKA_figures}
\end{figure}

Referring back to the example of LHC injection losses (Paragraph~\ref{par:LHC_450GeV_example}), both the radiation source and the geometry are very basic: the first consists of $450$~GeV protons travelling along the z-axis, and the second consists of a uniform aluminium target. In general, however, MC codes provide far more sophisticated source definitions, which users can adapt to their specific needs. Likewise, the geometrical description can be made substantially more complex, as illustrated by the detailed model of a section of the LHC tunnel shown in Fig.~\ref{fig:FLUKA_tunnel_geometry}. 

The propagation of particles in matter is described by a transport equation, which is solved numerically using the Monte Carlo method. For each particle, the mean free path is evaluated based on its type and its relevant interaction mechanisms in the material where it is located. A random step length to the next interaction is then sampled, followed by the selection of the interaction type according to the relative probabilities of the available channels. The final state of the interaction is generated, including the production of secondary particles when applicable, and the resulting energy deposition or other quantities are accumulated in statistical estimators of the desired observables. This procedure is repeated for the primary particle and all secondaries, as schematically shown in Fig.~\ref{fig:FLUKA_random_path}, until the particle history terminates, for example when the energy falls below predefined thresholds.

In all Monte Carlo simulations, the primary goal is to predict the values of physical observables. These predictions are obtained through a variety of statistical estimators, commonly referred to as scorings, which users select when configuring the simulation. Using FLUKA as an example (though similar concepts exist in other codes), scorings can predict the values of quantities such as energy deposition and its derivatives (e.g., dose), particle fluence or current as a function of energy, angle, or other kinematic variables, displacement per atom (DPA), residual activation, information on particle timing, and more. Scorings can be evaluated within specific regions, across material boundaries, or on region-independent spatial grids, and they can be recorded at the end of each simulation cycle or for individual events. Observables most relevant to particle accelerator applications will be discussed in greater detail in the next lecture of this series.

\section{A closer look at LHC-type radiation showers}
\label{sec:LHC_showers}

\begin{figure}[ht]
    \centering 
    \subfloat[\label{fig:FLUKA_protons_on_al_4ns}]{
    \includegraphics[width=0.328\linewidth]{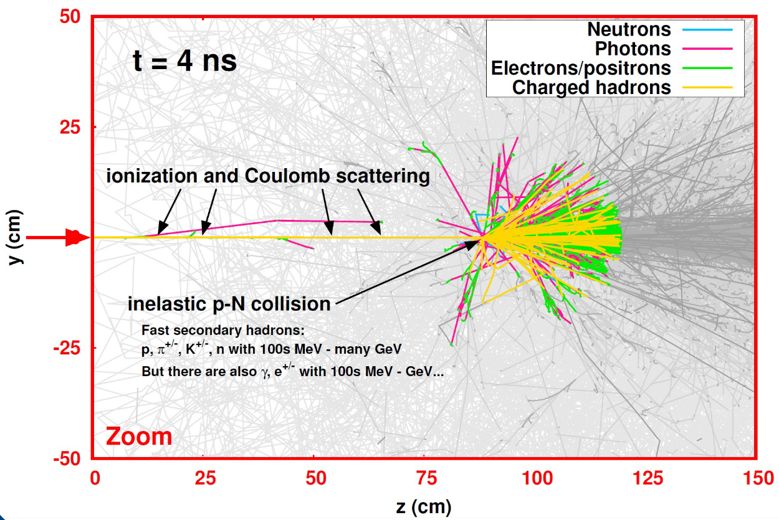}}    \subfloat[\label{fig:FLUKA_protons_on_al_12ns}]{
    \includegraphics[width=0.328\linewidth]{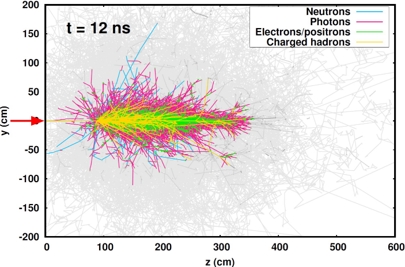}}
    \subfloat[\label{fig:FLUKA_protons_on_al_20ms}]{
    \includegraphics[width=0.328\linewidth]{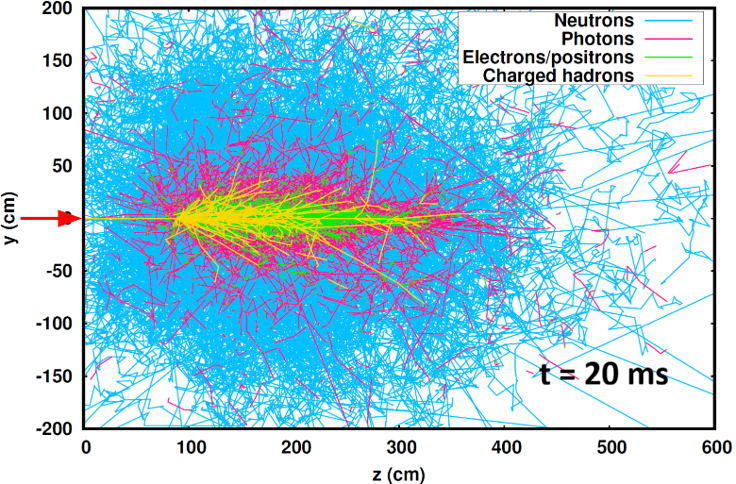}}
    \caption{FLUKA simulations of the products of $450$~GeV proton interactions in a uniform aluminium target, showing the particle tracks after (a) $4$~ns (zoomed over a smaller area), (b) $12$~ns, and (c) $20$~ms.  }
    \label{fig:FLUKA_450GeV_protons_on_aluminum_example}
\end{figure}

Having covered the key phenomenology of particle interactions with matter, it is instructive to examine a FLUKA simulation of a single $450$~GeV proton interacting with a uniform aluminium target, with the~generated particle tracks at different time delays illustrated in Fig.~\ref{fig:FLUKA_450GeV_protons_on_aluminum_example}. The first snapshot, taken $4$~ns after the proton enters the material along the z-axis, shows the primary proton track in yellow. Along its path, the proton experiences ionisation losses (as well as negligible Coulomb scattering) producing a few electrons ($\delta$-rays) that in turn generate secondary photons, and it eventually undergoes an inelastic collision with a nucleus, releasing a high multiplicity of secondary hadrons, as well as numerous photons and electrons/positrons. In the second snapshot, taken at $12$~ns over a larger volume, a hadronic shower is developing, embedding a significant EM component. Finally, at a much longer timescale of $20$~ms, a large number of neutron tracks can be observed extending far beyond the rest of the cascade. This behaviour arises because neutrons are long-lived compared to this timescale, are uncharged (and thus unaffected by ionisation losses), and interact primarily via elastic or inelastic collisions with nuclei. At distances far from the original inelastic collision, one can also observe occasional photons (and, more rarely, other hadrons) produced by neutron capture reactions.

\section{Summary and next steps}
\label{sec:summary}

This lecture reviewed the main mechanisms of particle-matter interactions, focusing on the fundamental concepts and emphasizing the aspects that are most relevant for the phenomenology of beam losses at high-energy accelerators. After introducing the interactions of photons and charged particles with matter, as well as nuclear reactions, particular attention was devoted to the properties of electromagnetic and hadronic showers, which develop in virtually all beam loss scenarios. The lecture then introduced the principles of Monte Carlo methods for particle–matter interaction simulations, using FLUKA to analyze the key features of an LHC-type radiation shower in matter. This lecture constitutes the first of a two-lecture series, with the second part~\cite{bib:LernerCAS2025beamlossconsequences} addressing the consequences of beam losses in high-energy accelerators.

\end{document}